\documentclass[showpacs,pra,twocolumn]{revtex4}
\usepackage{dcolumn}    
\usepackage{bm}         
\usepackage{ifpdf}
\usepackage{amssymb,lineno,amsfonts}
\usepackage{graphicx}   
\usepackage{bbm}
\usepackage{mathrsfs}
\usepackage{upgreek}
\usepackage{epstopdf}
\usepackage{setspace}
\usepackage{hyperref}
\usepackage[matrix,frame,arrow]{xypic}
\usepackage{rotating}
\usepackage{float}
\usepackage{natbib}
\usepackage{multirow}

\newcommand{\ket}[1]{\vert{#1}\rangle}
\newcommand{\bra}[1]{\langle{#1}\vert}
\newcommand{\outpr}[2]{\vert{#1}\rangle\langle{#2}\vert}
\newcommand{\inpr}[2]{\langle{#1}\vert{#2}\rangle}
\newcommand{\expv}[1]{\langle{#1}\rangle}

\newcommand{\proj}[1]{\outpr{#1}{#1}}

\begin{document}
\title{NMR Investigation of the Quantum Pigeonhole Effect}
\author{Anjusha V. S., Swathi S. Hegde, and T. S. Mahesh}
\email{mahesh.ts@iiserpune.ac.in}
\affiliation{Department of Physics and NMR Research Center,\\
Indian Institute of Science Education and Research, Pune 411008, India}

\begin{abstract}
{
NMR quantum simulators have been used for studying various quantum phenomena. Here, using a four-qubit NMR quantum simulator, we investigate the recently postulated quantum pigeonhole effect. In this phenomenon, a set of three particles in a two-path interferometer often appears to be in such a superposition that no two particles can be assigned a single path, thus exhibiting the non-classical behavior. In our experiments, quantum pigeons are emulated by three nuclear qubits whose states are probed jointly and noninvasively by an ancillary spin. The experimental results are in good agreement with quantum theoretical predictions.

}
\end{abstract}

\maketitle

\section{Introduction}
Quantum theory is well known for generating scenarios having no
classical analog. 
 A recent addition to this list is quantum pigeonhole effect (QPHE) \cite{aharonov2014quantum,aharonov2015weak}. In mathematics, the pigeonhole principle states that if $n$ items are
put in $m < n$  containers, then at least one container must have more than one item
\cite{milner1965pigeon}. While the method of contradiction is sufficient for proving the pigeonhole principle for a finite
number of objects, it calls for more advanced methods in the case of an infinite set of objects and containers.  The generalized
concept has several interesting consequences in mathematics  \cite{ru2001nevanlinna,buss1987polynomial,paris1988provability},
computer science \cite{ajtai1988complexity,factor2000caching,pape1982pigeon}, graph theory \cite{west2001introduction}, and combinatorics \cite{brualdi1992introductory}.

Recently, Aharonov and coworkers have theoretically illustrated 
certain quantum mechanical scenarios appearing to contradict the pigeonhole principle \cite{aharonov2014quantum}. 
This phenomenon, known as QPHE has already raised considerable interest.  For example 
Yu and Oh demonstrated 
the emergence of QPHE from quantum contextuality \cite{yu2014quantum}.  Rae and Forgan suggested that QPHE arises as a result of interference between the wavefunctions of weakly-interacting particles \cite{rae2014implications}.
In this work, we simulate QPHE using a four-qubit NMR quantum simulator.

Molecular nuclear spins in liquid ensembles controlled by NMR techniques 
have  successfully been
employed for simulating various quantum phenomena. 
They include simulating quantum harmonic oscillators \cite{somaroo1999quantum},
quantum phase transitions \cite{peng2005quantum},
ground state of hydrogen atom \cite{du2010nmr}, 
probabilities of particle in a potential \cite{shankar2014quantum}, and
quantum delayed choice experiment \cite{roy2012nmr}.
In this work we utilize a 4-qubit quantum register based on an ensemble of four mutually interacting spin-1/2 nuclei
partially oriented in a liquid crystal matrix.  While the rapid
isotropic reorientation of molecules in a liquid erases all the anisotropic interactions, only weak
indirect spin-spin interactions (J-couplings)
survive resulting in long nonlocal quantum gates that are comparable to decoherence time-scales.  On the other hand, in solid-state NMR systems, the strong intra- and intermolecular interactions inhibit high-fidelity quantum control of desired qubits.  In this sense, the partially oriented spin systems with no translational order have a specific advantage, i.e., negligible intermolecular interactions and residual intramolecular anisotropic interactions just strong enough for quantum control \cite{mahesh2006quantum}.
Accordingly, such systems are
increasingly being used for realizing quantum testbeds 
\cite{shankar2014quantum,rao2014efficient,du2010nmr}.

In the following section we provide a brief theoretical description of QPHE. In section III, we shall describe its NMR quantum simulation and finally we conclude in section IV.

\begin{figure}
	\includegraphics[trim=2cm 3.6cm 2cm 1cm, clip=true,width=8.5cm]{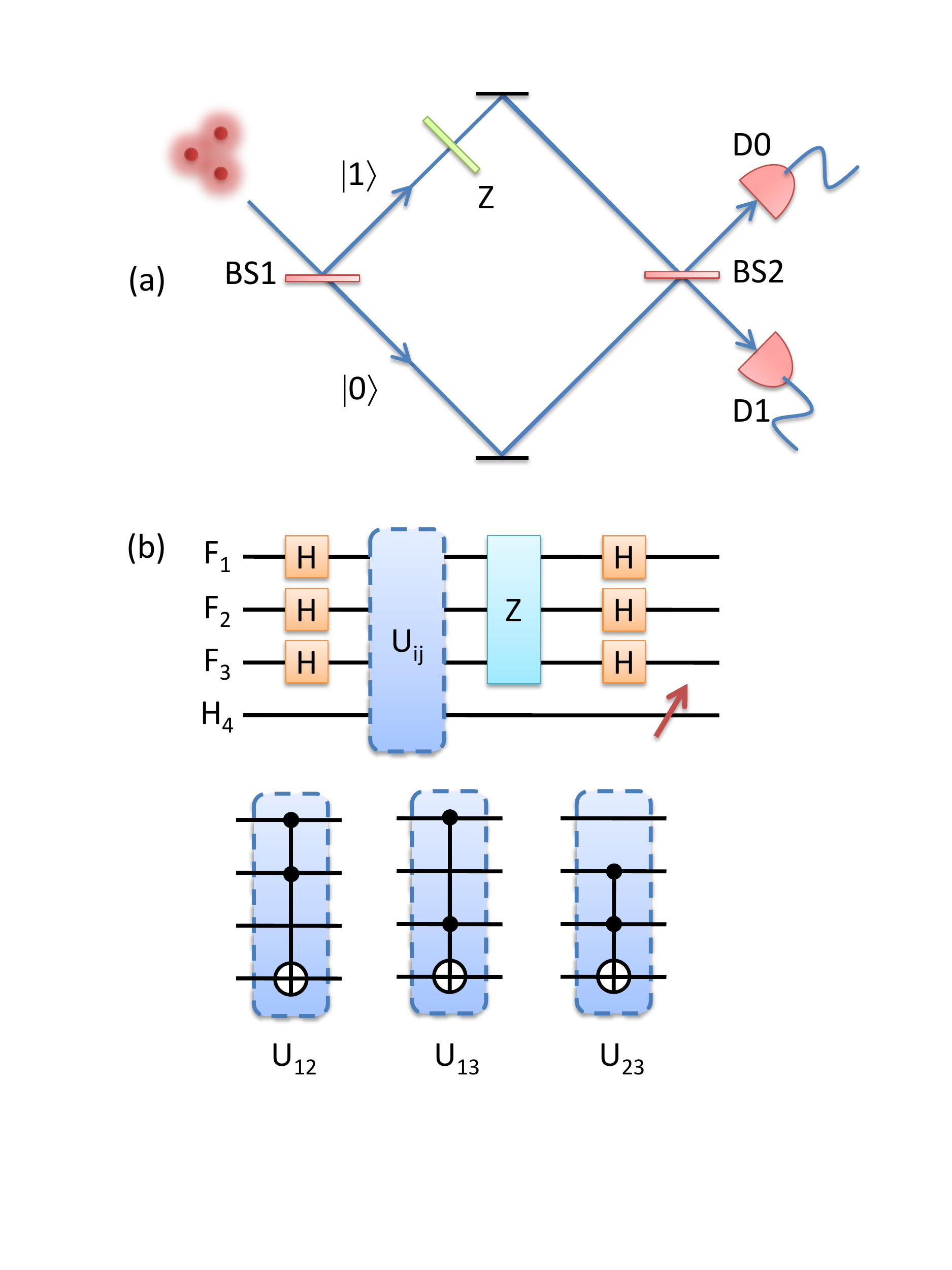}
	\caption{(a) Three quantum particles entering a Mach-Zehnder interferometer
		consisting of two beam-splitters (BS1 and BS2) and phase shifter (Z), and
		two particle-detectors D0 and D1.  (b) Circuit for NMR investigation of QPHE.  Hadamard
		gates perform the function of beam splitters, and Z-gate performs phase shift.
		Intermediate state information of the particle-qubits (F$_1$, F$_2$, F$_3$) is encoded onto an ancilla
		qubit (H$_4$) using one of the controlled operations $U_{12}$, $U_{13}$, and $U_{23}$.  The ancilla qubit is measured at the end of the circuit.
	}
	\label{mzi}
\end{figure}

\section{Theory}
Let us first consider a single quantum particle entering the Mach-Zehnder interferometer shown in Fig. \ref{mzi} (a).
It involves two beam-splitters (BS1 and BS2), a 90 degree phase-shifter (Z), and two
detectors (D0 and D1).  
BS1 is used to create a superposition of two paths labelled $\ket{0}$ and $\ket{1}$. 
When the particle initially prepared in state $\ket{0}$ enters BS1, it transforms to
$\ket{+} = (\ket{0} + \ket{1})/\sqrt{2}$.  Both paths are guided towards BS2 using mirrors.  After the phase-shifter Z, the state of the particle is
$\ket{+i} = (\ket{0} + i\ket{1})/\sqrt{2}$, and after BS2 it becomes 
$\{(1+i)\ket{0} + (1-i)\ket{1}\}/2$.  Thus the particle has equal probability of reaching either of the detectors.  
The state $\ket{+}$ can also be written in terms of $\ket{\pm i}$, i.e.,
\begin{eqnarray}
\ket{+} = \frac{1-i}{2}\ket{+i}+\frac{1+i}{2}\ket{-i}.
\end{eqnarray}
We notice that the first component, namely $\ket{+i}$ transforms to $\ket{-} = (\ket{0} - \ket{1})/\sqrt{2}$ after Z, and then to $\ket{1}$ after BS2, and finally 
ends up in detector D1.  Similarly, the second component, 
namely $\ket{-i}$ transforms to $\ket{+} = (\ket{0} + \ket{1})/\sqrt{2}$ after Z, and then to $\ket{0}$ after BS2, and finally ends up in detector D0.  In this sense,
a measurement outcome of $\ket{0}$ (or $\ket{1}$) amounts to postselecting $\ket{-i}$ (or $\ket{+i}$) state just before the phase-shifter.  

Let us now imagine the case of three particles in state $\ket{000}$ entering the interferometer.
After BS1, the state of the particles is
described by the superposition
$\ket{\psi_a} = \ket{+,+,+}$.
  The final state may collapse with equal probability to any one of the states $\{\ket{000},\ket{001},\ket{010},\ket{011},\ket{100},\ket{101},\ket{110},\ket{111} \}$.

The projectors
\begin{eqnarray}
P_{12}  = \proj{0} \otimes \proj{0} \otimes \mathbbm{1} + \proj{1} \otimes \proj{1} \otimes \mathbbm{1} \nonumber \\
P_{13}  = \proj{0} \otimes \mathbbm{1} \otimes \proj{0} + \proj{1} \otimes \mathbbm{1} \otimes \proj{1} \nonumber \\
P_{23}  = \mathbbm{1} \otimes \proj{0} \otimes \proj{0}   + \mathbbm{1} \otimes \proj{1} \otimes \proj{1},
\label{projs}
\end{eqnarray}
probe if any two of the particles are in the same state, i.e., $\ket{00}$ or $\ket{11}$.  The expectation values of the projectors give corresponding probabilities.  Evaluating the expectation values for the state $\ket{+,+,+}$, we find that $\expv{P_{12}} = \expv{P_{23}} = \expv{P_{13}} = 1/2$.  Just after BS1, the probability for any two particles being in the same path is therefore $1/2$.

We shall now consider only the cases wherein all the three particles reach the same detector, say D0 (or D1).
Then the measurement outcome $\ket{000}$ (or $\ket{111}$) is equivalent to postselecting the state $\ket{\phi_1} = \ket{-i,-i,-i}$ (or $\ket{\phi_0} =  \ket{+i,+i,+i}$) before the phase-shifter.

The projection $\ket{\psi_{j,k}^{same}} = P_{jk}\ket{\psi_a}$
describes the
component of $\ket{\psi_a}$ corresponding to particles $j$ and $k$ being in the same path.
Since,
\begin{eqnarray}
\inpr{\phi_0}{\psi_{1,2}^{same}} &=&
\bra{-i,-i,-i} {P_{12}}
\ket{+,+,+}
\nonumber \\
&=& 
\frac{\inpr{-i,-i,-i}{0,0,+}
+ \inpr{-i,-i,-i}{1,1,+}}
{2} \nonumber \\
& = & 0,
\end{eqnarray}
we conclude that the postselected state $\ket{\phi_0}$ has no component having particles 1 and 2 in the same path. 
Owing to the symmetry
in the pre- ($\ket{\psi_a}$) 
and post- ($\ket{\phi_0}$)
selected states, the above
conclusion can be extended to any pair of particles. This effect is interpreted as - if all the three particles have to reach the same detector D0, then no two particles can take the same path.  Of course, similar interpretation can also be given for the case in which all the three particles reach the same detector D1.  This phenomenon which seems to violate the classical pigeonhole effect is called QPHE.

\begin{figure}
	\includegraphics[trim=2cm 5cm 2cm 2cm, clip=true, width=8.5cm]{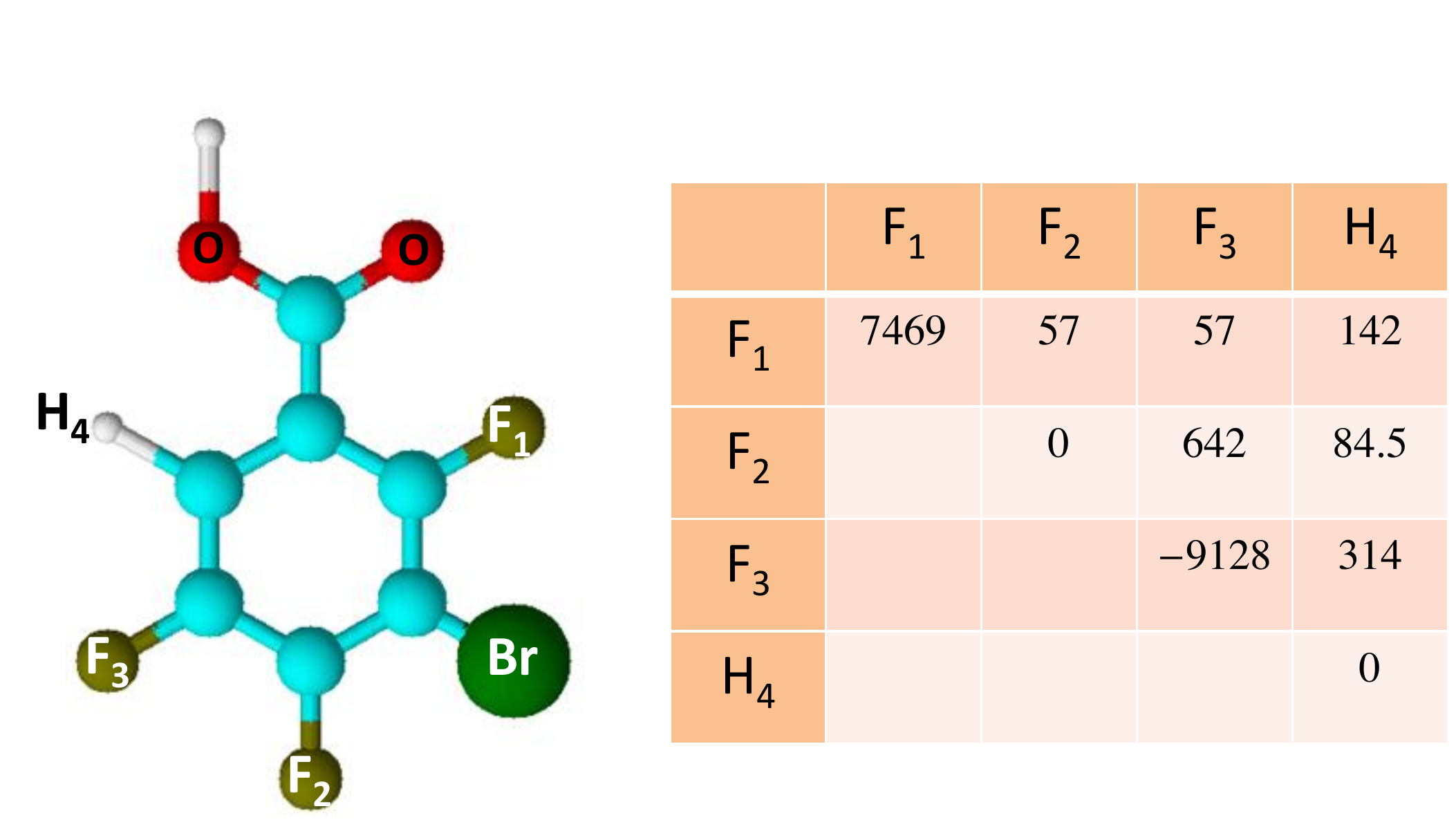}
	\caption{The molecular structure of 3-bromo-2,4,5-trifluorobenzoic acid. The chemical shifts (diagonal elements) and effective coupling constants $J_{ij}'$ (off-diagonal elements)  are shown in table.  The NMR spectrum of ancilla is shown in Fig. \ref{results}(a).}
	\label{spinsystem}
\end{figure}

\begin{figure}[b]
	\includegraphics [trim=3cm 4.5cm 2cm 4.5cm, clip=true,width=9cm]{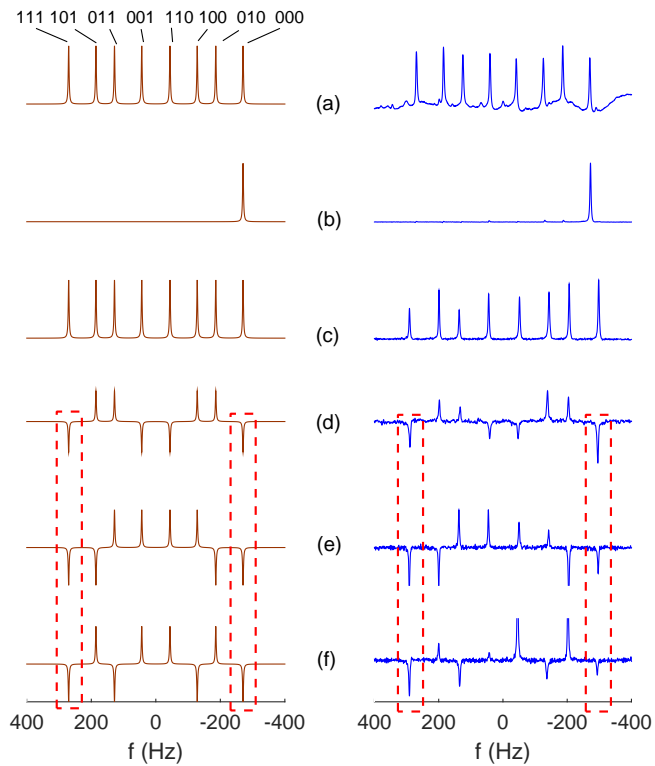}
	\caption{The ancilla NMR
		at various stages of QPHE simulation, each obtained by a final 90$^\circ$ detection pulse.  The simulated and experimental spectra are shown in the left and right columns respectively.
		(a) Thermal equilibrium state $\rho_{eq}$ (the background baseline is due to the liquid crystal signal),
		(b) The partial pseudopure state $\rho(0)$, and
		(c) after the entire MZI-circuit (without  $U_{ij}$) indicating the various combinations of detections.
		Spectra in (d-f) correspond to the complete QPHE circuit shown in Fig. \ref{mzi}(b) obtained with $U_{12}$, $U_{13}$, and $U_{23}$ respectively.
		The dashed boxes highlight the peaks corresponding to the postselected states as described in the text.}
	\label{results}
\end{figure}

\section{\label{sec:level2}NMR simulation}
The 4-qubit quantum register consists of 3-bromo-2,4,5-trifluorobenzoic acid (see Fig. \ref{spinsystem}) partially oriented in a liquid crystal N-(4-methoxybenzaldehyde)-4-butylanline (MBBA). The three $^{19}$F spins and 
the $^{1}$H spin are respectively used as the three particle qubits and the ancilla qubit.
All the experiments were carried out on a 500 MHZ Bruker NMR spectrometer at an ambient temperature of 298 K. The effective couplings in this system are  due to scalar interactions ($J_{ij}$) as well as partially averaged dipolar interactions ($D_{ij}$).
Thus the internal Hamiltonian
for the system, under weak-coupling approximation, can be written as
\begin{eqnarray}
H=-2\pi\sum\limits_{i}\nu_iI_{zi}+2\pi\sum_{ij}J_{ij}'I_{zi}I_{zj},
\end{eqnarray}
where $\nu_i$ are the resonance frequencies, $J_{ij}' = (J_{ij}+2D_{ij})$ are the effective coupling constants, and $I_{zi}$ is the z-component of spin angular momentum operator of $i$th spin
\cite{levitt2001spin}.
All the relevant Hamiltonian parameters like chemical shifts and effective coupling constants are tabulated in Fig. \ref{spinsystem}.

The first step in the experiment is the initialization of the 4-qubit system in a partial pseudopure state
\begin{equation}
\rho(0)= \mathbbm{1}/16
+\epsilon \ket{000}\bra{000} \otimes \sigma_z/2
\end{equation} 
where $\epsilon \sim 10^{-5}$ is the purity factor.  This state can be very easily realized
by taking difference between two initial states: (i) $\rho_{eq}$ corresponding to thermal equilibrium state and (ii) $\rho_{in}$ obtained by inverting the populations of levels $\ket{0000}$ and
$\ket{0001}$ using a low-power transition-selective Gaussian pulse of duration 80 ms \cite{fung2001pairs}.  The ancilla spectrum corresponding to the initial state obtained by a
$90^\circ$ detection pulse (shown in Fig. \ref{results} (b))
clearly indicates the transition corresponding to the particle-state $\ket{000}$.

\begin{figure}
	\includegraphics [trim=2.8cm 0cm 1.5cm 1cm, clip=true,width=9cm]{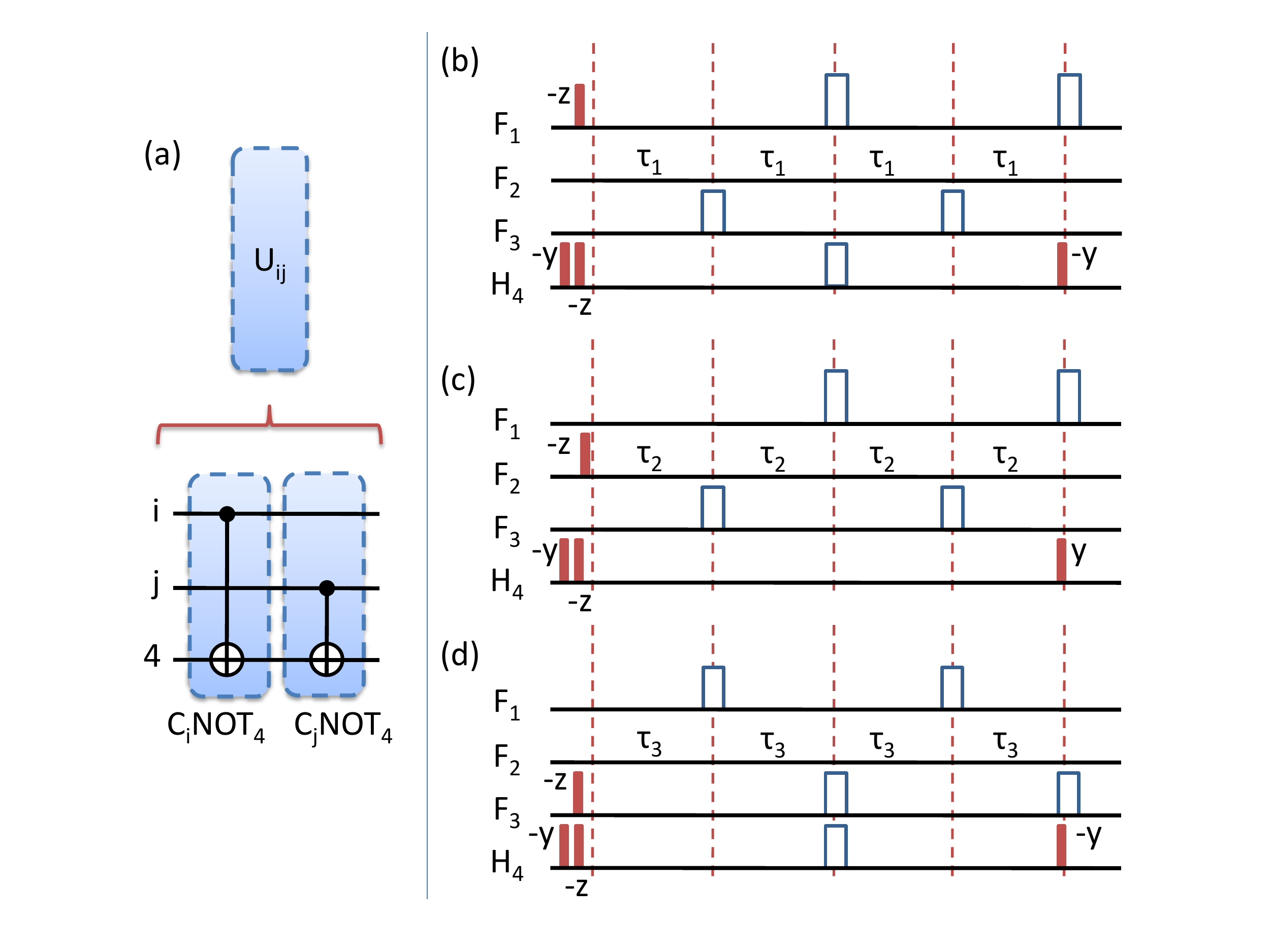}
	\caption{
		(a) Realizing $U_{ij}$ by a pair of CNOT gates. (b-d) NMR pulse sequences corresponding to C$_1$NOT$_4$, C$_2$NOT$_4$, and C$_3$NOT$_4$ respectively.  
		All the $\pi$ pulses (open rectangles) are about $y$ axis and the phases of the $\pi/2$ pulses (filled rectangles) are as indicated.  The delays
		are set according to $\tau_{i}$ = $1/(8J_{i,4}')$.}
	\label{cnots}
\end{figure}

The next step is to apply the various elements of the Mach-Zehnder interferometer as shown in Fig. \ref{mzi} (b).  All the unitary operations were realized by GRAPE optimal control technique  \cite{khaneja2005optimal}.
The durations of the GRAPE pulses ranged from 400 $\upmu$s to 700 $\upmu$s, and the fidelities were  better than 0.99 over a 10 \% RF inhomogeneity range.

  The experimental spectrum after the complete MZI-circuit is shown in Fig. \ref{results}(c).
To probe the component of the wave-function prior to the phase gate, we insert operator $U_{ij}$ as shown in Fig. \ref{mzi} (b).  
The action of $U_{ij}$ is
given by
\begin{eqnarray}
U_{ij} = P_{ij} \otimes \mathbbm{1}_a
+(\mathbbm{1}_s-P_{ij})
\otimes X_a,
\end{eqnarray}
where $P_{ij}$ are the projections as described in eqns. \ref{projs},
$\mathbbm{1}_s$, $\mathbbm{1}_a$ are respectively the identity operators on system and ancilla, and 
$X_a$ is the NOT operator on ancilla.
Effectively, $U_{ij}$ conserves the ancilla if particles $i$ and $j$ are in the same path, but inverts it otherwise. Thus in the ancilla NMR spectrum, a positive ancilla peak indicates two particles being in the same path, while a negative peak indicates them being in different paths.  Each of the $U_{ij}$ operations is realized by a pair of CNOT gates as described in Fig. \ref{cnots} (a).
The NMR pulse-sequences to generate each of the CNOT gates are shown in Fig. \ref{cnots} (b-d).

The NMR spectra obtained
after the complete MZI-circuit along with probing of particles $(1,2)$ (by $U_{12}$), $(1,3)$ (by $U_{13}$), and $(2,3)$ (by $U_{23}$), are shown respectively on the right side of Figs. \ref{results}(d-f). 
The corresponding simulated spectra are shown on the left side. The intensity variations in the experimental spectra are mainly due to the imperfections in executing CNOT operations resulting from RF field inhomogeneities, nonlinearities of the RF amplifiers, as well as decoherence.
However, we observe an overall agreement of the experimental results with the quantum theoretical simulations.
In particular, we focus on the NMR transition 000 (111) corresponding to postselection of all the three particles reaching D0 (D1). These transitions are highlighted by dashed lines in Fig. \ref{results}(d-f)).
All these transitions have negative intensities indicating that no two particles are in the same path.  This is a clear demonstration of QPHE.  

For three particles and two containers, the various other arrangements and their classical as well as quantum possibilities are shown in Table 1.
Here the first two arrangements, where there is a clear contradiction between classical and quantum, correspond to QPHE.  For the other cases, while the possibilities are probabilistic in the classical regime, they are certain in the quantum regime.  Of course, it is also possible to generalize QPHE to $N$ particles in $M<N$ containers \cite{aharonov2014quantum}.\\

\begin{table}
	\begin{center}
	\includegraphics [trim=1cm 9.6cm 0.5cm 1cm, clip=true,width=8.5cm]{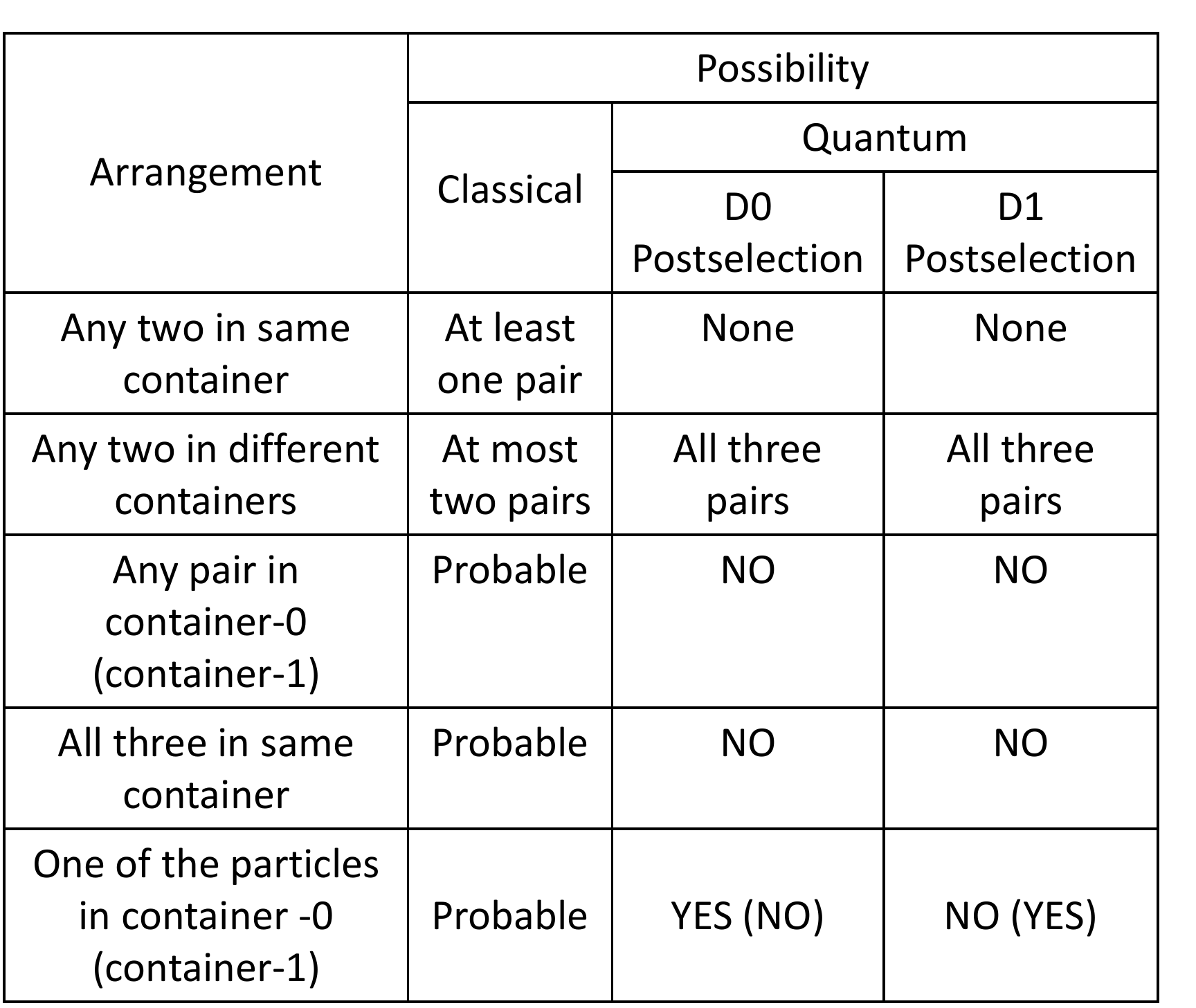}
	\caption{The classical and quantum possibilities are tabulated for various arrangements of three particles in two containers.  The top two rows correspond to QPHE.}
	\end{center}
\end{table}

\section{Conclusions}
Quantum pigeonhole effect is yet another illustration of quantum systems displaying effects beyond the classical predictions. 
The nonclassical effect in this case emerges as a result of assignment of premeasurement states based on the knowledge of postmeasurement values.  

 Here we provided the first experimental simulation of   QPHE using an NMR quantum simulator.  The quantum register consisted of three $^{19}$F spins simulating three quantum particles whose intermediate state was probed by a $^{1}$H spin (ancilla). 
The experimental results agreed well with the quantum theoretical predictions. 
The successful demonstration of QPHE also illustrated good quantum control achieved on a four-qubit heteronuclear NMR system partially oriented in a liquid crystal.

\section*{Acknowledgements}
Authors thank Sharad Joshi, Abhishek Shukla, Sudheer Kumar, and Deepak Khurana for valuable discussions and IISER, Pune, for the funding.

\bibliographystyle{apsrev4-1}
\bibliography{bib2}

\begin{thebibliography}{23}%
\makeatletter
\providecommand \@ifxundefined [1]{%
 \@ifx{#1\undefined}
}%
\providecommand \@ifnum [1]{%
 \ifnum #1\expandafter \@firstoftwo
 \else \expandafter \@secondoftwo
 \fi
}%
\providecommand \@ifx [1]{%
 \ifx #1\expandafter \@firstoftwo
 \else \expandafter \@secondoftwo
 \fi
}%
\providecommand \natexlab [1]{#1}%
\providecommand \enquote  [1]{``#1''}%
\providecommand \bibnamefont  [1]{#1}%
\providecommand \bibfnamefont [1]{#1}%
\providecommand \citenamefont [1]{#1}%
\providecommand \href@noop [0]{\@secondoftwo}%
\providecommand \href [0]{\begingroup \@sanitize@url \@href}%
\providecommand \@href[1]{\@@startlink{#1}\@@href}%
\providecommand \@@href[1]{\endgroup#1\@@endlink}%
\providecommand \@sanitize@url [0]{\catcode `\\12\catcode `\$12\catcode
  `\&12\catcode `\#12\catcode `\^12\catcode `\_12\catcode `\%12\relax}%
\providecommand \@@startlink[1]{}%
\providecommand \@@endlink[0]{}%
\providecommand \url  [0]{\begingroup\@sanitize@url \@url }%
\providecommand \@url [1]{\endgroup\@href {#1}{\urlprefix }}%
\providecommand \urlprefix  [0]{URL }%
\providecommand \Eprint [0]{\href }%
\providecommand \doibase [0]{http://dx.doi.org/}%
\providecommand \selectlanguage [0]{\@gobble}%
\providecommand \bibinfo  [0]{\@secondoftwo}%
\providecommand \bibfield  [0]{\@secondoftwo}%
\providecommand \translation [1]{[#1]}%
\providecommand \BibitemOpen [0]{}%
\providecommand \bibitemStop [0]{}%
\providecommand \bibitemNoStop [0]{.\EOS\space}%
\providecommand \EOS [0]{\spacefactor3000\relax}%
\providecommand \BibitemShut  [1]{\csname bibitem#1\endcsname}%
\let\auto@bib@innerbib\@empty
\bibitem [{\citenamefont {Aharonov}\ \emph {et~al.}(2014)\citenamefont
  {Aharonov}, \citenamefont {Colombo}, \citenamefont {Popescu}, \citenamefont
  {Sabadini}, \citenamefont {Struppa},\ and\ \citenamefont
  {Tollaksen}}]{aharonov2014quantum}%
  \BibitemOpen
  \bibfield  {author} {\bibinfo {author} {\bibfnamefont {Y.}~\bibnamefont
  {Aharonov}}, \bibinfo {author} {\bibfnamefont {F.}~\bibnamefont {Colombo}},
  \bibinfo {author} {\bibfnamefont {S.}~\bibnamefont {Popescu}}, \bibinfo
  {author} {\bibfnamefont {I.}~\bibnamefont {Sabadini}}, \bibinfo {author}
  {\bibfnamefont {D.~C.}\ \bibnamefont {Struppa}}, \ and\ \bibinfo {author}
  {\bibfnamefont {J.}~\bibnamefont {Tollaksen}},\ }\href@noop {} {\bibfield
  {journal} {\bibinfo  {journal} {arXiv preprint arXiv:1407.3194}\ } (\bibinfo
  {year} {2014})}\BibitemShut {NoStop}%
\bibitem [{\citenamefont {Aharonov}\ and\ \citenamefont
  {Cohen}(2015)}]{aharonov2015weak}%
  \BibitemOpen
  \bibfield  {author} {\bibinfo {author} {\bibfnamefont {Y.}~\bibnamefont
  {Aharonov}}\ and\ \bibinfo {author} {\bibfnamefont {E.}~\bibnamefont
  {Cohen}},\ }\href@noop {} {\bibfield  {journal} {\bibinfo  {journal} {arXiv
  preprint arXiv:1504.03797}\ } (\bibinfo {year} {2015})}\BibitemShut {NoStop}%
\bibitem [{\citenamefont {Milner}\ and\ \citenamefont
  {Rado}(1965)}]{milner1965pigeon}%
  \BibitemOpen
  \bibfield  {author} {\bibinfo {author} {\bibfnamefont {E.~C.}\ \bibnamefont
  {Milner}}\ and\ \bibinfo {author} {\bibfnamefont {R.}~\bibnamefont {Rado}},\
  }\href@noop {} {\bibfield  {journal} {\bibinfo  {journal} {Proceedings of the
  London Mathematical Society}\ }\textbf {\bibinfo {volume} {3}},\ \bibinfo
  {pages} {750} (\bibinfo {year} {1965})}\BibitemShut {NoStop}%
\bibitem [{\citenamefont {Ru}(2001)}]{ru2001nevanlinna}%
  \BibitemOpen
  \bibfield  {author} {\bibinfo {author} {\bibfnamefont {M.}~\bibnamefont
  {Ru}},\ }\href@noop {} {\emph {\bibinfo {title} {Nevanlinna theory and its
  relation to Diophantine approximation}}}\ (\bibinfo  {publisher} {World
  Scientific},\ \bibinfo {year} {2001})\BibitemShut {NoStop}%
\bibitem [{\citenamefont {Buss}(1987)}]{buss1987polynomial}%
  \BibitemOpen
  \bibfield  {author} {\bibinfo {author} {\bibfnamefont {S.~R.}\ \bibnamefont
  {Buss}},\ }\href@noop {} {\bibfield  {journal} {\bibinfo  {journal} {The
  Journal of Symbolic Logic}\ }\textbf {\bibinfo {volume} {52}},\ \bibinfo
  {pages} {916} (\bibinfo {year} {1987})}\BibitemShut {NoStop}%
\bibitem [{\citenamefont {Paris}\ \emph {et~al.}(1988)\citenamefont {Paris},
  \citenamefont {Wilkie},\ and\ \citenamefont {Woods}}]{paris1988provability}%
  \BibitemOpen
  \bibfield  {author} {\bibinfo {author} {\bibfnamefont {J.~B.}\ \bibnamefont
  {Paris}}, \bibinfo {author} {\bibfnamefont {A.~J.}\ \bibnamefont {Wilkie}}, \
  and\ \bibinfo {author} {\bibfnamefont {A.~R.}\ \bibnamefont {Woods}},\
  }\href@noop {} {\bibfield  {journal} {\bibinfo  {journal} {The Journal of
  Symbolic Logic}\ }\textbf {\bibinfo {volume} {53}},\ \bibinfo {pages} {1235}
  (\bibinfo {year} {1988})}\BibitemShut {NoStop}%
\bibitem [{\citenamefont {Ajtai}(1988)}]{ajtai1988complexity}%
  \BibitemOpen
  \bibfield  {author} {\bibinfo {author} {\bibfnamefont {M.}~\bibnamefont
  {Ajtai}},\ }in\ \href@noop {} {\emph {\bibinfo {booktitle} {Foundations of
  Computer Science, 1988., 29th Annual Symposium on}}}\ (\bibinfo
  {organization} {IEEE},\ \bibinfo {year} {1988})\ pp.\ \bibinfo {pages}
  {346--355}\BibitemShut {NoStop}%
\bibitem [{\citenamefont {Factor}\ and\ \citenamefont
  {Farchi}(2000)}]{factor2000caching}%
  \BibitemOpen
  \bibfield  {author} {\bibinfo {author} {\bibfnamefont {M.~E.}\ \bibnamefont
  {Factor}}\ and\ \bibinfo {author} {\bibfnamefont {E.~D.}\ \bibnamefont
  {Farchi}},\ }\href@noop {} {\enquote {\bibinfo {title} {Caching in a data
  processing system using the pigeon hole principle},}\ } (\bibinfo {year}
  {2000}),\ \bibinfo {note} {uS Patent 6,094,706}\BibitemShut {NoStop}%
\bibitem [{\citenamefont {Pape}\ \emph {et~al.}(1982)\citenamefont {Pape},
  \citenamefont {Riepe}, \citenamefont {Schopper} \emph
  {et~al.}}]{pape1982pigeon}%
  \BibitemOpen
  \bibfield  {author} {\bibinfo {author} {\bibfnamefont {H.}~\bibnamefont
  {Pape}}, \bibinfo {author} {\bibfnamefont {L.}~\bibnamefont {Riepe}},
  \bibinfo {author} {\bibfnamefont {J.}~\bibnamefont {Schopper}},  \emph
  {et~al.},\ }\href@noop {} {\bibfield  {journal} {\bibinfo  {journal} {The Log
  Analyst}\ }\textbf {\bibinfo {volume} {23}} (\bibinfo {year}
  {1982})}\BibitemShut {NoStop}%
\bibitem [{\citenamefont {West}\ \emph {et~al.}(2001)\citenamefont {West} \emph
  {et~al.}}]{west2001introduction}%
  \BibitemOpen
  \bibfield  {author} {\bibinfo {author} {\bibfnamefont {D.~B.}\ \bibnamefont
  {West}} \emph {et~al.},\ }\href@noop {} {\emph {\bibinfo {title}
  {Introduction to graph theory}}},\ Vol.~\bibinfo {volume} {2}\ (\bibinfo
  {publisher} {Prentice hall Upper Saddle River},\ \bibinfo {year}
  {2001})\BibitemShut {NoStop}%
\bibitem [{\citenamefont {Brualdi}(1992)}]{brualdi1992introductory}%
  \BibitemOpen
  \bibfield  {author} {\bibinfo {author} {\bibfnamefont {R.~A.}\ \bibnamefont
  {Brualdi}},\ }\href@noop {} {\emph {\bibinfo {title} {Introductory
  combinatorics}}}\ (\bibinfo  {publisher} {New York},\ \bibinfo {year}
  {1992})\BibitemShut {NoStop}%
\bibitem [{\citenamefont {Yu}\ and\ \citenamefont {Oh}(2014)}]{yu2014quantum}%
  \BibitemOpen
  \bibfield  {author} {\bibinfo {author} {\bibfnamefont {S.}~\bibnamefont
  {Yu}}\ and\ \bibinfo {author} {\bibfnamefont {C.}~\bibnamefont {Oh}},\
  }\href@noop {} {\bibfield  {journal} {\bibinfo  {journal} {arXiv preprint
  arXiv:1408.2477}\ } (\bibinfo {year} {2014})}\BibitemShut {NoStop}%
\bibitem [{\citenamefont {Rae}\ and\ \citenamefont
  {Forgan}(2014)}]{rae2014implications}%
  \BibitemOpen
  \bibfield  {author} {\bibinfo {author} {\bibfnamefont {A.}~\bibnamefont
  {Rae}}\ and\ \bibinfo {author} {\bibfnamefont {T.}~\bibnamefont {Forgan}},\
  }\href@noop {} {\bibfield  {journal} {\bibinfo  {journal} {arXiv preprint
  arXiv:1412.1333}\ } (\bibinfo {year} {2014})}\BibitemShut {NoStop}%
\bibitem [{\citenamefont {Somaroo}\ \emph {et~al.}(1999)\citenamefont
  {Somaroo}, \citenamefont {Tseng}, \citenamefont {Havel}, \citenamefont
  {Laflamme},\ and\ \citenamefont {Cory}}]{somaroo1999quantum}%
  \BibitemOpen
  \bibfield  {author} {\bibinfo {author} {\bibfnamefont {S.}~\bibnamefont
  {Somaroo}}, \bibinfo {author} {\bibfnamefont {C.}~\bibnamefont {Tseng}},
  \bibinfo {author} {\bibfnamefont {T.}~\bibnamefont {Havel}}, \bibinfo
  {author} {\bibfnamefont {R.}~\bibnamefont {Laflamme}}, \ and\ \bibinfo
  {author} {\bibfnamefont {D.~G.}\ \bibnamefont {Cory}},\ }\href@noop {}
  {\bibfield  {journal} {\bibinfo  {journal} {Physical review letters}\
  }\textbf {\bibinfo {volume} {82}},\ \bibinfo {pages} {5381} (\bibinfo {year}
  {1999})}\BibitemShut {NoStop}%
\bibitem [{\citenamefont {Peng}\ \emph {et~al.}(2005)\citenamefont {Peng},
  \citenamefont {Du},\ and\ \citenamefont {Suter}}]{peng2005quantum}%
  \BibitemOpen
  \bibfield  {author} {\bibinfo {author} {\bibfnamefont {X.}~\bibnamefont
  {Peng}}, \bibinfo {author} {\bibfnamefont {J.}~\bibnamefont {Du}}, \ and\
  \bibinfo {author} {\bibfnamefont {D.}~\bibnamefont {Suter}},\ }\href@noop {}
  {\bibfield  {journal} {\bibinfo  {journal} {Physical Review A}\ }\textbf
  {\bibinfo {volume} {71}},\ \bibinfo {pages} {012307} (\bibinfo {year}
  {2005})}\BibitemShut {NoStop}%
\bibitem [{\citenamefont {Du}\ \emph {et~al.}(2010)\citenamefont {Du},
  \citenamefont {Xu}, \citenamefont {Peng}, \citenamefont {Wang}, \citenamefont
  {Wu},\ and\ \citenamefont {Lu}}]{du2010nmr}%
  \BibitemOpen
  \bibfield  {author} {\bibinfo {author} {\bibfnamefont {J.}~\bibnamefont
  {Du}}, \bibinfo {author} {\bibfnamefont {N.}~\bibnamefont {Xu}}, \bibinfo
  {author} {\bibfnamefont {X.}~\bibnamefont {Peng}}, \bibinfo {author}
  {\bibfnamefont {P.}~\bibnamefont {Wang}}, \bibinfo {author} {\bibfnamefont
  {S.}~\bibnamefont {Wu}}, \ and\ \bibinfo {author} {\bibfnamefont
  {D.}~\bibnamefont {Lu}},\ }\href@noop {} {\bibfield  {journal} {\bibinfo
  {journal} {Physical review letters}\ }\textbf {\bibinfo {volume} {104}},\
  \bibinfo {pages} {030502} (\bibinfo {year} {2010})}\BibitemShut {NoStop}%
\bibitem [{\citenamefont {Shankar}\ \emph {et~al.}(2014)\citenamefont
  {Shankar}, \citenamefont {Hegde},\ and\ \citenamefont
  {Mahesh}}]{shankar2014quantum}%
  \BibitemOpen
  \bibfield  {author} {\bibinfo {author} {\bibfnamefont {R.}~\bibnamefont
  {Shankar}}, \bibinfo {author} {\bibfnamefont {S.~S.}\ \bibnamefont {Hegde}},
  \ and\ \bibinfo {author} {\bibfnamefont {T.}~\bibnamefont {Mahesh}},\
  }\href@noop {} {\bibfield  {journal} {\bibinfo  {journal} {Physics Letters
  A}\ }\textbf {\bibinfo {volume} {378}},\ \bibinfo {pages} {10} (\bibinfo
  {year} {2014})}\BibitemShut {NoStop}%
\bibitem [{\citenamefont {Roy}\ \emph {et~al.}(2012)\citenamefont {Roy},
  \citenamefont {Shukla},\ and\ \citenamefont {Mahesh}}]{roy2012nmr}%
  \BibitemOpen
  \bibfield  {author} {\bibinfo {author} {\bibfnamefont {S.~S.}\ \bibnamefont
  {Roy}}, \bibinfo {author} {\bibfnamefont {A.}~\bibnamefont {Shukla}}, \ and\
  \bibinfo {author} {\bibfnamefont {T.}~\bibnamefont {Mahesh}},\ }\href@noop {}
  {\bibfield  {journal} {\bibinfo  {journal} {Physical Review A}\ }\textbf
  {\bibinfo {volume} {85}},\ \bibinfo {pages} {022109} (\bibinfo {year}
  {2012})}\BibitemShut {NoStop}%
\bibitem [{\citenamefont {Mahesh}\ and\ \citenamefont
  {Suter}(2006)}]{mahesh2006quantum}%
  \BibitemOpen
  \bibfield  {author} {\bibinfo {author} {\bibfnamefont {T.}~\bibnamefont
  {Mahesh}}\ and\ \bibinfo {author} {\bibfnamefont {D.}~\bibnamefont {Suter}},\
  }\href@noop {} {\bibfield  {journal} {\bibinfo  {journal} {Physical Review
  A}\ }\textbf {\bibinfo {volume} {74}},\ \bibinfo {pages} {062312} (\bibinfo
  {year} {2006})}\BibitemShut {NoStop}%
\bibitem [{\citenamefont {Rao}\ \emph {et~al.}(2014)\citenamefont {Rao},
  \citenamefont {Mahesh},\ and\ \citenamefont {Kumar}}]{rao2014efficient}%
  \BibitemOpen
  \bibfield  {author} {\bibinfo {author} {\bibfnamefont {K.~R.~K.}\
  \bibnamefont {Rao}}, \bibinfo {author} {\bibfnamefont {T.}~\bibnamefont
  {Mahesh}}, \ and\ \bibinfo {author} {\bibfnamefont {A.}~\bibnamefont
  {Kumar}},\ }\href@noop {} {\bibfield  {journal} {\bibinfo  {journal}
  {Physical Review A}\ }\textbf {\bibinfo {volume} {90}},\ \bibinfo {pages}
  {012306} (\bibinfo {year} {2014})}\BibitemShut {NoStop}%
\bibitem [{\citenamefont {Levitt}(2001)}]{levitt2001spin}%
  \BibitemOpen
  \bibfield  {author} {\bibinfo {author} {\bibfnamefont {M.~H.}\ \bibnamefont
  {Levitt}},\ }\href@noop {} {\emph {\bibinfo {title} {Spin dynamics: basics of
  nuclear magnetic resonance}}}\ (\bibinfo  {publisher} {John Wiley \& Sons},\
  \bibinfo {year} {2001})\BibitemShut {NoStop}%
\bibitem [{\citenamefont {Fung}(2001)}]{fung2001pairs}%
  \BibitemOpen
  \bibfield  {author} {\bibinfo {author} {\bibfnamefont {B.}~\bibnamefont
  {Fung}},\ }\href@noop {} {\bibfield  {journal} {\bibinfo  {journal} {The
  Journal of Chemical Physics}\ }\textbf {\bibinfo {volume} {115}},\ \bibinfo
  {pages} {8044} (\bibinfo {year} {2001})}\BibitemShut {NoStop}%
\bibitem [{\citenamefont {Khaneja}\ \emph {et~al.}(2005)\citenamefont
  {Khaneja}, \citenamefont {Reiss}, \citenamefont {Kehlet}, \citenamefont
  {Schulte-Herbr{\"u}ggen},\ and\ \citenamefont {Glaser}}]{khaneja2005optimal}%
  \BibitemOpen
  \bibfield  {author} {\bibinfo {author} {\bibfnamefont {N.}~\bibnamefont
  {Khaneja}}, \bibinfo {author} {\bibfnamefont {T.}~\bibnamefont {Reiss}},
  \bibinfo {author} {\bibfnamefont {C.}~\bibnamefont {Kehlet}}, \bibinfo
  {author} {\bibfnamefont {T.}~\bibnamefont {Schulte-Herbr{\"u}ggen}}, \ and\
  \bibinfo {author} {\bibfnamefont {S.~J.}\ \bibnamefont {Glaser}},\
  }\href@noop {} {\bibfield  {journal} {\bibinfo  {journal} {Journal of
  Magnetic Resonance}\ }\textbf {\bibinfo {volume} {172}},\ \bibinfo {pages}
  {296} (\bibinfo {year} {2005})}\BibitemShut {NoStop}%
\end{thebibliography}%

\end{document}